\documentstyle[11pt]{article}
\newcommand{\be}{\begin{equation}}
\newcommand{\ee}{\end{equation}}
\newcommand{\bea}{\begin{eqnarray}}
\newcommand{\eea}{\end{eqnarray}}
\newcommand{\nn}{\nonumber}
\def\a{\alpha}
\def\b{\beta}
\def \eps {\epsilon}
\def\s{\sigma}
\def\d{\delta}

\def\t{\tau}
\def\cf{{\cal F}}

\begin{document}
\title{ Comment on "Constraint Quantization of Open String in Background
B field and Noncommutative D-brane" }
\author{F. Loran\thanks{e-mail:
loran@cc.iut.ac.ir}\\ \\
  {\it Department of  Physics, Isfahan University of Technology (IUT)}\\
{\it Isfahan,  Iran,} \\
  {\it Institute for Studies in Theoretical Physics and Mathematics (IPM)}\\
{\it P. O. Box: 19395-5531, Tehran, Iran.}}
\date{}
\maketitle

\begin{abstract}
In the paper "Constraint Quantization of Open String in Background $B$ field and
Noncommutative D-brane", it is claimed that the boundary conditions lead to an
infinite set of secondary constraints and Dirac brackets result in a non-commutative
Poisson structure for D-brain. Here we show that contrary to the arguments in that
paper, the set of secondary constraints on the boundary is finite and the
non-commutativity algebra can not be obtained by evaluating the Dirac brackets.
\end{abstract}
In ref. \cite{Ho}, Chong-Sun Chu and Pei-Ming Ho have studied the constraint
quantization of open string in background $B$ field. They have obtained an infinite
set of secondary constraints due to the boundary conditions for the open string on
D-brane. Then, they have shown that the Dirac brackets result in the
non-commutativity algebra derived in ref. \cite{Ho1} for the end points of the open
string and consequently the D-brane becomes non-commutative. This result is very
interesting because, as far as we know, this is one of the most important
applications of the Dirac method of quantization of the secondary constraints
\cite{Dirac}. Here, we show that the Dirac method does not lead to an infinite set of
secondary constraints but to a set of finite second class constraint chains which
does not lead to the non-commutativity algebra.
\par
The action for an open string ending on a D$p$-brane is \cite{Ho, Ho1} \be S_B=
\frac{1}{4\pi\a'} \int_{\Sigma} d^2\sigma \left[ g^{\a\b}G_{\mu\nu} \partial_\a
X^{\mu}\partial_\b X^{\nu}+
\cf_{ij}\partial_{\a}X^i\partial_{\b}X^j\right],\label{p1}\ee where \be \cf=B-dA=B-F,
\ee is the modified Born-Infeld field strength. The equation of motion is \be
(\partial^2_{\t}-\partial^2_{\s}) X^\mu =0, \ee and the boundary conditions at $\s =0,
\pi$ are: \bea &\partial_\s X^i + \partial_\t X^j \cf_j{}^i =0, \quad i,j= 0,1,\cdots,
p,
\label{a2}\\
&X^a =x_0^a,\quad a =p+1, \cdots, D.  \eea In the following, for simplicity and
without loss of generality, we assume the case $p=D$. Defining the momentum fields \be
\Pi_i(\t,\s)=\frac{\d}{\d X^i(\s,\t)}S_B=\frac{1}{2\pi\a'}\left(\partial_\t X_i +
\partial_\s X^j\cf_{ji}\right),
\ee  then the Hamiltonian is \be H=\frac{1}{4\pi\a'}\int
d\s\left[\left(2\pi\a'\Pi-\partial_\s X.\cf\right)^2+ \left(\partial_\s
X\right)^2\right]+\lambda^0_i \Phi^i_0+\lambda^\pi_i\Phi^i_\pi, \ee where
$M_{ij}=\eta_{ij}-\cf_i{}^k\cf_{kj}$, $\lambda_i^\s$'s are Lagrange multipliers and
$\Phi^i_\s$'s are the primary constraints corresponding to the boundary conditions
given in Eq.(\ref{a2}) \cite{Ho,She}: \be \label{a3} \Phi^i_\s=\int d\s'
\d(\s'-\s)\phi(\s'),\hspace{1cm}\s=0,\pi,\ee in which \be \phi(\s)=2\pi\a'
\Pi^k(\s)\cf_k{}^i +\partial_{\s}X^j(\s)M_j{}^i. \ee The secondary constraints can be
obtained by considering the consistency conditions of the primary constraints: \be
\dot{\Phi}^i_\s=0\to\Psi^i_\s=\int d\s'\d(\s-\s')\psi(\s')=0,\hspace{1cm}\s=0,\pi,\ee
where \be \psi(\s)=
\partial_{\s}\Pi(\s).\ee This result is the direct consequence of the fact that
(see Eq.(28) in ref.\cite{Ho}), \bea\frac{1}{2\pi\a'}
\{\phi^i(\t,\s),\phi^j(\t,\s')\}&=&-\partial_{\s'}\d(\s-\s')
\cf_k{}^iM_{k'}{}^j\eta^{kk'}+\partial_\s
\d(\s-\s')M_{k}{}^i\cf_{k'}{}^j\eta^{kk'}\nn\\
&=&\partial_\s\d(\s-\s')\left(\cf^{ki}M_k{}^j+M^{ki}\cf_k{}^j\right)\nn\\
&=&\partial_\s\d(\s-\s')\left(-\cf M+M\cf\right)^{ij}\nn\\&=&0,\label{a4} \eea and
consequently, \be \{\Phi^i_\s,\Phi^j_{\s'}\}=0,\hspace{1cm}\s,\s'=0,\pi.\ee
\footnote{It is worth noting that if we had
$\det\left(\{\Phi^i_\s,\Phi^j_{\s'}\}\right)\neq0,$ then no secondary constraint
should be introduced, since the consistency conditions $\dot{\Phi}^i_\s=0$ would
determine the Lagrange multipliers \cite{Dirac}. }To obtain the final result given in
Eq.(\ref{a4}) we have used the following properties: \bea
\{X^i(\t,\s),\Pi^j(\t,\s')\}&=&\d(\s-\s')\eta^{ij},\nn\\
\left(\cf\right)^{ij}&=&-\left(\cf\right)^{ji},\nn\\
\left(M\right)^{ij}&=&+\left(M\right)^{ji}.\eea Since
\be\{\phi^i(\t,\s),\psi^j(\t,\s')\}=M^{ij}
\partial_\s\partial_{\s'}\d(\s-\s')\neq 0,\label{a5}\ee the constraints $\Phi^i_\s$'s
and $\Psi^i_\s$'s form a set of secondary constraints \cite{Ho,She}. Consequently the
consistency of the secondary constraints $\Psi^i_\s$'s determine the Lagrange
multiplier and according to the well known arguments in the context of constrained
systems, no additional constraint emerges. To calculate the Dirac brackets, it is
suitable to define constraints \be\Omega^a_\s=\int d\s'
\d(\s-\s')\omega^a(\s'),\hspace{1cm} a=1,\cdots,2D,\ee for $ \s=0,\pi,$ where
$\omega^a$'s are defined as follows: \bea
\omega^i&=&\phi^i,\nn\\\omega^{D+i}&=&\psi^i,\hspace{1cm}i=1,\cdots,D.\eea The matrix
of the Poisson brackets of the constraints $\Omega^a_0$'s is \be
C=\left(\begin{array}{cc}0&M\\-M&0\end{array}\right) \int d\s d\s'\d(\s)\d(\s')
\partial_\s\partial_{\s'}\d(\s-\s'). \ee  Using the
equality \be \delta(\s-\s')=\lim_{\eps\rightarrow 0}\;\; {1\over \eps\sqrt\pi}
e^{{-(\s-\s')^2 \over \eps^2}},\ee the inverse of the matrix $C$ can be obtained as
follows: \be C^{-1}= \lim_{\eps\rightarrow
0}\frac{\epsilon^3\sqrt{\pi}}{2}\left(\begin{array}{cc}0&-M^{-1}\\M^{-1}&0
\end{array}\right).\ee By definition, the Dirac bracket
of the fields $X^i(\t,0)$ is:
 \bea&&\{X^i(\t,0),X^j(\t,0)\}_{DB}=\nn\\&-&\int d\s
 d\s'\left(\sum_{i,j=1}^2\d(\s-\s_i)\d(\s'-\s_j)\right)
 \{X^i(\t,0),\omega^a(\t,\s)\}C^{-1}_{ab}\{\omega^b(\t,\s'),X^j(\t,0)\}\nn\\
 &\sim& \int d\s d\s'\left(\sum_{i,j=1}^2\d(\s-\s_i)\d(\s'-\s_j)\right)
 \d(\s)\partial_{\s'}\d(\s')\nn\\&=&0,
 \label{b}\eea
 where $\s_1=0$ and $\s_2=\pi$.\par
 Consequently the Dirac
method of constraint quantization leads to a commutative Poisson structure for
$D$-branes to which the open string end points are attached \cite{He}. Finally it is
necessary to note that the final result given in Eq.(\ref{b}) does not change if one
insists on the assertion that an infinite set of constraints exist on the boundaries
\cite{She}.

\end{document}